\documentclass[a4paper,11pt]{article}
\pdfoutput=1

\usepackage{jheppub}

\usepackage[utf8]{inputenc}
\usepackage[T1]{fontenc}
\usepackage{lmodern, amsmath, amssymb, slashed, graphicx, comment,adjustbox,bm}
\usepackage[dvipsnames]{xcolor}
\usepackage{soul}
\usepackage{url}
\usepackage{multirow}
\usepackage{array}
\usepackage{booktabs}
\usepackage{floatrow}
\newfloatcommand{capbtabbox}{table}[][\FBwidth]
\usepackage{braket}

\def\figureautorefname~#1\null{Fig.\,#1\null}

\def\equationautorefname~#1\null{Eq.\,(#1)\null}

\newcommand{\A}{\mathcal{A}}

\newcommand{\bpm}{\begin{pmatrix}}
\newcommand{\epm}{\end{pmatrix}}

\newcommand{\beq}{\begin{equation} }
\newcommand{\eeq}{\end{equation} }

\title{Positivity Bounds in Scalar Effective Field Theories at One-loop Level}

\author[a]{Yunxiao Ye,}
\author[a]{Bin He,}
\author[a,b]{Jiayin Gu}

\affiliation[a]{Department of Physics and Center for Field Theory and Particle Physics, Fudan University, 2005 Song Hu Road, Shanghai 200438, China}
\affiliation[b]{Key Laboratory of Nuclear Physics and Ion-beam Application (MOE), Fudan University, 220 Handan Road, Shanghai 200433, China}

\emailAdd{yxye22@m.fudan.edu.cn}
\emailAdd{bhe22@m.fudan.edu.cn}
\emailAdd{jiayin\_gu@fudan.edu.cn}

\abstract{  
Parameters in an effective field theory can be subject to certain positivity bounds if one requires a UV completion that obeys the fundamental principles of quantum field theory.  These bounds are relatively straightforward at the tree level, but would become more obscure when loop effects are important.  Using scalar theories as examples, we carefully exam the positivity bounds in a case where the leading contribution to a forward elastic amplitude arises at the one-loop level, and point out certain subtleties in terms of the implications of positivity bounds on the theory parameter space.  In particular, the one-loop generated dimension-8 operator coefficients (that would be positive if generated at the tree level), as well as their $\beta$-functions are generally not subject to positivity bounds as they might correspond to interference terms of the cross sections under the optical theorem, which could have either sign.  A strict positivity bound can only be implied when all contributions at the same loop order are considered, including the ones from dim-4 and dim-6 operator coefficients, which have important effects at the one-loop level.  Our results may have important implications on the robustness of experimental tests of positivity bounds.  
}

\begin{document} 
\maketitle
\flushbottom

\section{Introduction}
\label{Introduction}

Effective Field Theory (EFT) is a useful framework that connects the physics at different scales.  In a bottom-up approach, the low-energy effects of the   UV physics can be parameterized by a series of higher-dimensional operators, suppressed by a cutoff scale $\Lambda$.  If the UV physics is unknown, the coefficients of these operators, known as Wilson coefficients, should be treated as free parameters to be measured by experiments.  Given the large number of parameters, it is desirable to understand how physical principles can reduce the freedom in this huge landscape of parameters. 
It is well known that a certain set of dimension-8 (dim-8) Wilson coefficients are subject to a class of constraints, known as positivity bounds, derived from the fundamental principles of quantum field theory (QFT) including unitarity, analyticity, Lorentz invariance and crossing symmetry~\cite{Adams:2006sv,Distler:2006if,Manohar:2008tc,Nicolis:2009qm,Bellazzini:2014waa,Bellazzini:2016xrt,deRham:2017avq,deRham:2017zjm,Bellazzini:2017fep,deRham:2017xox,Bellazzini:2017bkb,deRham:2018qqo,Bellazzini:2019bzh,Wang:2020jxr,Zhang:2020jyn,Alberte:2020jsk,Tokuda:2020mlf,Bellazzini:2020cot,Tolley:2020gtv,Trott:2020ebl,Herrero-Valea:2020wxz,Sinha:2020win,Arkani-Hamed:2020blm,Caron-Huot:2020cmc,Alberte:2020bdz,Li:2021lpe,Davighi:2021osh,Alberte:2021dnj,Bellazzini:2021oaj,Chiang:2021ziz,Bern:2021ppb,Guerrieri:2021tak,Caron-Huot:2021rmr,Du:2021byy,Aoki:2021ckh,Henriksson:2021ymi,Chiang:2022ltp,deRham:2022hpx,Chiang:2022jep,Caron-Huot:2022ugt,Riembau:2022yse,Fernandez:2022kzi,CarrilloGonzalez:2022fwg,Hong:2023zgm,CarrilloGonzalez:2023cbf,Hamada:2023cyt,Bellazzini:2023nqj, Bittar:2024xuc}.  
Important applications have been found in the Standard Model Effective Field Theory (SMEFT)~\cite{Low:2009di,Bellazzini:2018paj,Zhang:2018shp,Bi:2019phv, Remmen:2019cyz,Remmen:2020vts,Fuks:2020ujk,Yamashita:2020gtt,Gu:2020ldn,Bonnefoy:2020yee,Remmen:2020uze,Gu:2020thj,Chala:2021wpj, Zhang:2021eeo,Azatov:2021ygj,Li:2022tcz, Li:2022rag, Li:2022aby,Ghosh:2022qqq,Remmen:2022orj,Chen:2023bhu, Gu:2023emi, Davighi:2023acq,  Chala:2023jyx, Chala:2023xjy, Altmannshofer:2023bfk, DasBakshi:2023htx}.  One important aspect of positivity bounds in the SMEFT is that in some cases these bounds can be explicitly tested by experiments (see {\it e.g.} Ref.~\cite{Gu:2020ldn,Gu:2023emi}), which in principle provides a test on the fundamental principles of QFT.  

The implications of positivity bounds are most straightforward at the tree level, where the 4-point amplitudes can be written as polynomials of the Mandelstam variables.  The situation is more complicated at the loop level, as the loop contributions generally introduce logarithmic dependence of the Mandelstam variables.  The renormalization group (RG) evolution of the dim-8 Wilson coefficients may also have important effects.  In some cases, the loop effects have been found to have important impacts on the positivity bounds.  
Ref.~\cite{Bellazzini:2020cot, Bellazzini:2021oaj} discussed some important IR effects that can arise at the one-loop level.  
Ref.~\cite{Chala:2021wpj} pointed out an explicit example where a loop-generated dim-8 Wilson coefficient could violate the na\"ive tree-level positivity bound and also studied its RG running effects. 
Ref.~\cite{Li:2022aby} promoted the convex cone method in Ref.~\cite{Zhang:2020jyn} to the one-loop level and applied it to the Higgs sector in SMEFT.  
The impacts on the RG running were further examined in Ref.~\cite{Chala:2023jyx,Chala:2023xjy}, which claim that positivity bounds impose nontrivial constraints on the dim-8 anomalous dimension matrix and could determine the signs of some of its entries.   

\begin{figure}[t]
	\centering
	\includegraphics[width=\textwidth]{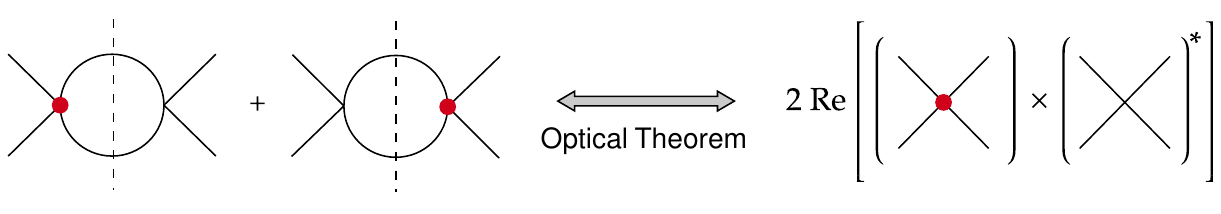}
	\caption{The one-loop diagram with one insertion of a dim-8 operator (indicated by the red dot) of the forward elastic amplitude corresponds to an interference term in the total cross section.  Note that the contact dim-8 interaction will be replaced by some renormalizable interactions (typically with a heavy particle propagator) in the UV.}
    \label{fig:optical}
\end{figure}

In this paper, we further investigate on the implications of positivity bounds at the one-loop level.  In stead of the full SMEFT (which is rather complicated), we focus on a simple EFT with two real scalars ($\phi_1, \, \phi_2$) with quartic couplings, which already have nontrivial structures at the one-loop level such as the RG mixing between different operators. 
To obtain a rigorous interpretation of the positivity bounds, we carefully compute all the contributions in the EFT, up to the one-loop and dim-8 level (in the loop and EFT expansions, respectively), to the dispersion relation (shown in \autoref{eq:Sigmapos}), and then apply the positivity bound to study the impacts on the Wilson coefficients.   
We also consider a toy UV model involving a heavy scalar $\Phi$ and a $\Phi\phi_1\phi_2$ trilinear coupling, in which the 4-point amplitude of $\phi_1\phi_1 \to \phi_1\phi_1$, which exhibits a positivity bound, can only be generated at the one-loop level.  This provides an explicit example to check the implications of positivity bounds.  
Our main finding is that at the one-loop level, the dim-4 and dim-6 operators also have important contributions to the dispersion relation, and if they are included,   
the RG mixing among different dim-8 operators should not be subject to positivity bounds.  
This can be understood from the optical theorem, which for a massless scalar $\phi$ states that 
\beq
\frac{1}{s} \, {\mathrm{Im}} \left[\A(\phi\phi \to \phi\phi)|_{t\to 0}\right] = \sigma(\phi \phi \to X) \,, 
\eeq
where $X$ denotes all possible final states.  The positivity bound on the Wilson coefficients is implied from the positivity of the total cross section $\sigma(\phi \phi \to X)$.  
However, as illustrated in \autoref{fig:optical}, the RG mixing diagrams correspond to the interference terms of the cross section 
under the optical theorem.  While the total cross section (which contains the dim-4 and dim-6 contributions) is positive, the interference terms could take either sign, so the positivity bound would not apply to the RG mixing diagrams alone.    
The same argument also applies to the one-loop generated dim-8 Wilson coefficients (that would be positive if generated at the tree level), 
and they are not necessarily subject to the same tree-level positivity bounds.  
Some of our results have been pointed out by previous studies, such as Ref.~\cite{Chala:2021wpj}, which also mentioned the one-loop contribution from dim-4 operators to the dispersion relation.  However, to the best of our knowledge, our study is the first that considers the full one-loop contribution to the dispersion relation, including not only the RG contributions but also the dim-8 one-loop finite terms.  Therefore, the positivity bounds we obtain are rigorous up to the one-loop and dim-8 level.  We also make the important connection between the apparent violation of positivity bounds and the contributions in the UV model from the view of the optical theorem, which was not discussed in the previous studies.




%

The rest of this paper is organized as follows: In \autoref{sec:model} we lay down the details of the models we study, including both the general 2-scalar  EFT in \autoref{sec:mdeft} and the specific UV model with a $\Phi\phi_1\phi_2$ trilinear coupling in \autoref{sec:mduv}.  
In \autoref{sec:pos}, we review the derivation of positivity bounds from dispersion relations, and apply the bound on the $\phi_1\phi_1 \to \phi_1\phi_1$ amplitude in the 2-scalar EFT in \autoref{sec:poseft}.  Then, in \autoref{sec:posuv}, we look at the dispersion relation and the positivity bound from the UV perspective in the $\Phi\phi_1\phi_2$ model, and verify that the bound is automatically satisfied as long as all the relevant contributions in the dispersion relation are included.     
Finally, we conclude in \autoref{sec:con}. 
The full one-loop $\beta$-functions and one-loop matching results are provided in \autoref{app:beta}.  
The results for the $\phi_1\phi_2 \to \phi_1\phi_2$ amplitude are provided in \autoref{app:A12}.


\section{The scalar model}
\label{sec:model}


\subsection{The 2-scalar EFT}
\label{sec:mdeft}


We focus on the EFT of two light real scalars $\phi_1$ and $\phi_2$ with the Lagrangian
\beq
\mathcal{L} = \dfrac{1}{2} \partial_\mu \phi_1 \partial^\mu \phi_1 + \dfrac{1}{2} \partial_\mu \phi_2 \partial^\mu \phi_2 - V(\phi_1, \phi_2) \, ,
\label{eq:Left} 
\eeq
where the potential $V(\phi_1, \phi_2)$ contains all possible interaction terms, including higher dimensional operators parameterized by $c^{(n)}_i \mathcal{O}^{(n)}_i / \Lambda^{n-4}$, where $n$ denotes the mass dimension of the operator and $\Lambda$ is the cutoff scale ({\it i.e.} the masses of the heavy particles in the UV theory).  Assuming $\Lambda$ is sufficiently large,\footnote{More explicitly, we require $s_0 \ll \Lambda^2$ where $s_0$ will be defined in \autoref{sec:pos}.} we could truncate the EFT series and keep only operators of dimension 8 or less.  For simplification, we also make a number of additional assumptions.  First, we take the masses of $\phi_1$ and $\phi_2$ to zero.  In general, this can be problematic for the dispersion relation, as the branch cut on the real axis of $s$ is extended down to zero and covers the whole real axis in this cases.  However, as mentioned later in \autoref{sec:pos}, it is possible to exploit the crossing symmetry of a real scalar and still obtain a meaningful dispersion relation in this case.   
We also impose separate $\mathbb{Z}_{2}$ symmetries on $\phi_1$ and $\phi_2$ which greatly reduces the number of possible interaction terms.  In particular, all trilinear couplings, as well as operators with odd mass dimensions, are forbidden.\footnote{As a result, there is no $t$-channel exchange of a massless particle, which ensures that the forward amplitude is finite, at least at tree level.}

We will focus on the operators that contribute to the tree-level 4-point  amplitudes, since only such operators contribute to the dispersion relation of a 4-point amplitude up to the one-loop level.\footnote{
Note also that operators of the form $\phi^6$ could generate a 4-point amplitude at one-loop but would not contribute to the dispersion relation since its loop is independent of the external momenta.} 
It is most convenient to parameterize these operators by tree-level on-shell amplitudes~\cite{Shadmi2019, Ma:2019gtx, Durieux:2019siw, Gu:2020thj, AccettulliHuber:2021uoa, DeAngelis:2022qco}, and the results are summarized in \autoref{tab:eft}.  
In particular, all 4-point amplitudes are generated by contact interactions (since there is no trilinear interaction) and are thus polynomials of the Mandelstam variables $s$, $t$ and $u$, defined as (assuming all momenta are outgoing)
\begin{equation}
    s = \left( p_1 + p_2 \right)^2, \quad t = \left( p_1 + p_3 \right)^2, \quad u = \left( p_1 + p_4 \right)^2 \,. 
\end{equation}
A 4-point amplitude is dimensionless, and can be written in the general form
\beq
\A = \sum_{n}  \A^{[n]}  = \sum_{n,i}  c^{[n]}_i A^{[n]}_i \,, \label{eq:ampcom}
\eeq
where $c^{[n]}_i$ is the Wilson coefficient (which absorbs the $1/\Lambda^{n-4}$ factor) of operator $\mathcal{O}^{(n)}_i$ with $n$ the mass dimension, and $A^{[n]}_i$ contain only the kinematic variables.  Therefore, $A^{[4]}_i$, $A^{[6]}_i$ and $A^{[8]}_i$ have mass dimensions 0, 2 and 4, respectively, and it is straightforward to enumerate all possible combinations of $s$, $t$ and $u$ for a fixed dimension.  The crossing symmetry of amplitudes and the (massless) relation $s+t+u=0$ imposes further restrictions on the form of the amplitudes, which results in the general parameterization in \autoref{tab:eft}.\footnote{For instance, $\A_1\equiv A(\phi_1\phi_1\phi_1\phi_1)$ is symmetrical in $s$, $t$ and $u$ since it is invariant under the crossing of any two external legs.  This restricts $\A^{[6]}_1 \propto s+t+u =0$.  See {\it e.g.} Refs.~\cite{Shadmi2019, Gu:2020thj} for more details.}
\begin{table}[tbp]
\centering
\begin{tabular}{cccc}
& $\A_1 \equiv \A(\phi_1\phi_1\phi_1\phi_1)$ & $\A_2 \equiv \A(\phi_2\phi_2\phi_2\phi_2)$ & $\A_{12} \equiv \A(\phi_1\phi_2\phi_1\phi_2)$  \\
\toprule 
 $\displaystyle D4$ & $\displaystyle \mathcal{A}_{1}^{[4]} = c_{1}^{[4]}$ & $\displaystyle \mathcal{A}_{2}^{[4]}= c_{2}^{[4]}$ & $\displaystyle \mathcal{A}_{12}^{[4]} = c_{12}^{[4]}$   \\
 \\
 $\displaystyle D6$ & $\displaystyle \mathcal{A}_{1}^{[6]} = 0 $ & $\displaystyle \mathcal{A}_{2}^{[6]} = 0 $ & $ \displaystyle \mathcal{A}_{12}^{[6]} =  c_{12}^{[6]}\, t $ \\
 \\
$\displaystyle D8$ & $\displaystyle \mathcal{A}_{1}^{[8]} = c_{1}^{[8]} (s^{2} + t^{2} + u^{2}) $  & $\displaystyle \mathcal{A}_{2}^{[8]} = c_{2}^{[8]} (s^{2} + t^{2} + u^{2})  $ & $\displaystyle \mathcal{A}_{12}^{[8]} =  c_{12,su}^{[8]} (s^{2} + u^{2}) + c_{12,t}^{[8]} t^{2} $ \\
 \bottomrule
\end{tabular}
\caption{Tree-level amplitude basis for 4-point amplitudes in the 2 scalar EFT.  Each amplitude is written in a general form involving the $s,t,u$ parameters that are consistent with dimensional analysis and crossing symmetry.  Each independent kinematic term has a free coefficient $c^{[n]}_i$ where $n$ is the corresponding operator dimension.  Note the momentum label of the external particles are always assigned in the order $\A(1234)$.}
\label{tab:eft}
\end{table}
It is straightforward to translate the amplitude basis in \autoref{tab:eft} to the Lagrangian in \autoref{eq:Left}, which can be written as
\begin{equation}
    \begin{split}
        \mathcal{L} = & \dfrac{1}{2} \partial_\mu \phi_1 \partial^\mu \phi_1 + \dfrac{1}{2} \partial_\mu \phi_2 \partial^\mu \phi_2 + \frac{ c_{1}^{[4]} }{ 4! } \phi_{1}^{4} + \frac{ c_{2}^{[4]} }{ 4! } \phi_{2}^{4} + \frac{ c_{12}^{[4]} }{ 4 } \phi_{1}^{2} \phi_{2}^{2} +  c_{12}^{[6]} \left( \partial_\mu \phi_1 \right) \left( \partial^\mu \phi_2 \right) \phi_1 \phi_2 \\
        &  + \frac{ c_{1}^{[8]} }{ 2 } \left( \partial_\mu \phi_1 \right) \left( \partial^\mu \phi_1 \right) \left( \partial_\nu \phi_1 \right) \left( \partial^\nu \phi_1 \right) + \frac{ c_{2}^{[8]} }{ 2 } \left( \partial_\mu \phi_2 \right) \left( \partial^\mu \phi_2 \right) \left( \partial_\nu \phi_2 \right) \left( \partial^\nu \phi_2 \right) \\
        &  + 2 c_{12,su}^{[8]} \left( \partial_\mu \phi_1 \right) \left( \partial^\mu \phi_2 \right) \left( \partial_\nu \phi_1 \right) \left( \partial^\nu \phi_2 \right) + c_{12,t}^{[8]} \left( \partial_\mu \phi_1 \right) \left( \partial^\mu \phi_1 \right) \left( \partial_\nu \phi_2 \right) \left( \partial^\nu \phi_2 \right) ,
    \end{split}
\end{equation}
where all the numerical factors of interaction terms are appropriately normalized such that the tree-level amplitudes coincide with \autoref{tab:eft}.
Finally, since our focus is on the one-loop amplitudes, the one-loop $\beta$-functions of the Wilson coefficients are particularly important for us. For the dim-8 coefficients, they are of the general form
\begin{equation}
    \beta_{i}^{[8]} \equiv \mu \frac{ d c_{i}^{[8]} }{ d \mu } = \frac{1}{16 \pi^{2}} \left( \gamma_{ij} c_{j}^{[8]} + \gamma_{ijk}^{\prime} c_{j}^{[6]} c_{k}^{[6]} \right) \,, \label{eq:beta}
\end{equation}
where the coefficients $\gamma_{ij}$ (which absorbs a factor of $c^{[4]}_k$) and $\gamma_{ijk}^{\prime}$ are conventionally denoted as the anomalous dimension matrices. The explicit results of the $\beta$-functions can be found in \autoref{app:beta}.  


\subsection{The $\Phi\phi_1\phi_2$ UV model}
\label{sec:mduv}

\begin{figure}[t]
    \centering
    \includegraphics[width=\textwidth]{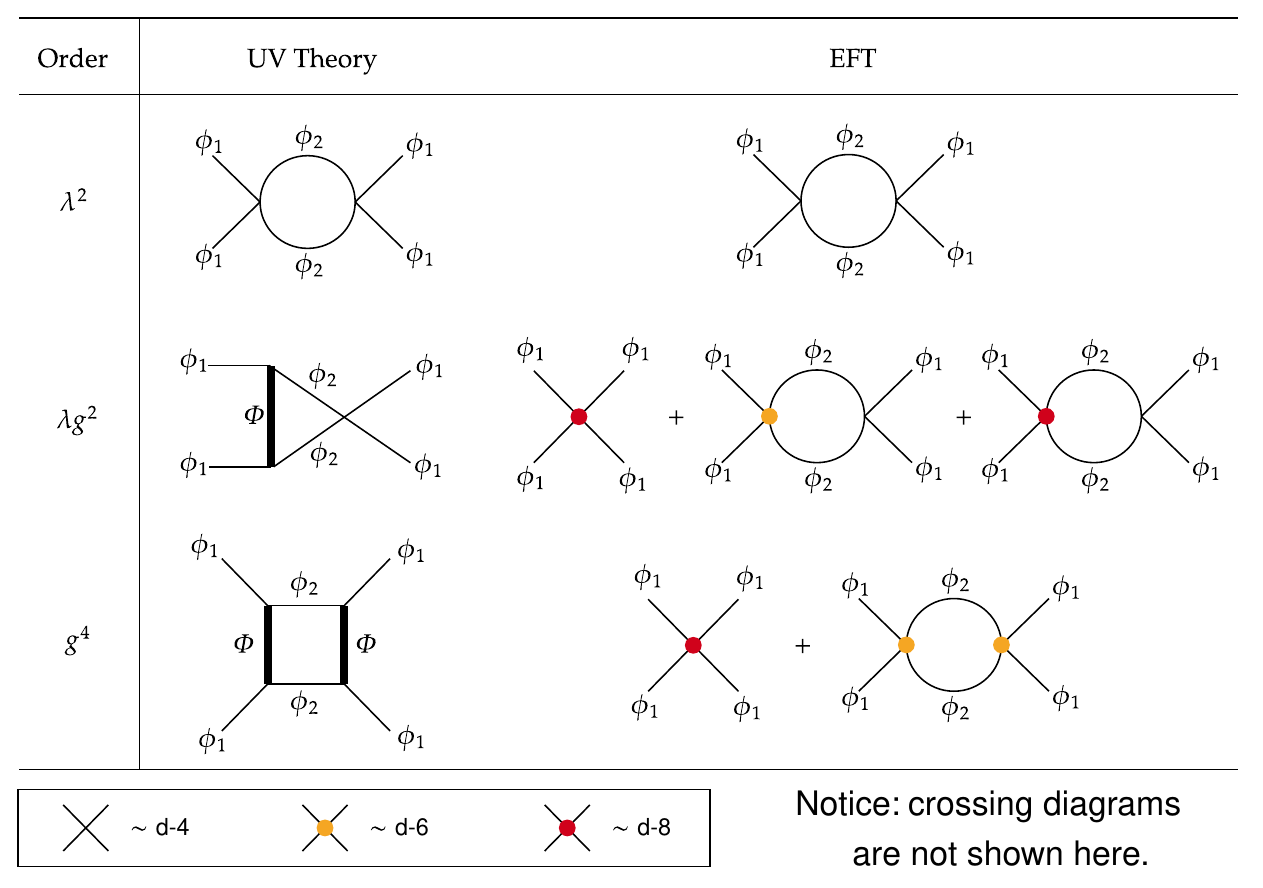}
    \caption{ One-loop
diagrams which contribute to $\A_1 \equiv \A(\phi_1 \phi_1 \phi_1 \phi_1)$ in the scalar theory of \autoref{eq:phiuv}.  The three types of diagrams are proportional to $\lambda^2$, $\lambda g^2$ and $g^4$, respectively.  Additional diagrams obtained by crossing the external $\phi_1$s are not explicitly shown here.
The EFT is obtained by integrating out the heavy scalar $\Phi$.  Note that there is also a $g^2$ contribution to the renormalizable coupling $c^{[4]}_{12}$ (see \autoref{eq:ctreemat}) which is not explicitly shown here. }
    \label{fig:aphi1}
\end{figure}

With the general 2-scalar EFT in the previous section, we shall also consider a specific UV model which involves a heavy scalar $\Phi$ with a $\Phi \phi_1\phi_2$ trilinear coupling.  We also include a $\phi_1^2 \phi_2^2$ quartic coupling, while all other couplings are set to zero.  Its Lagrangian is given by
\begin{equation}
    \mathcal{L} = \frac{1}{2} \left( \partial^{\mu} \Phi \partial_{\mu} \Phi - M^{2} \Phi^{2} \right) + \frac{1}{2} \partial^{\mu} \phi_{1} \partial_{\mu} \phi_{1} + \frac{1}{2} \partial^{\mu} \phi_{2} \partial_{\mu} \phi_{2} - g M \Phi \phi_{1} \phi_{2} - \frac{1}{4} \lambda \phi_{1}^{2} \phi_{2}^{2} \,, 
\label{eq:phiuv}
\end{equation}
where $M$ is the mass of $\Phi$, and $g$ is the trilinear coupling of $\Phi \phi_1\phi_2$, with the additional factor of $M$ to make $g$ dimensionless. The $\mathbb{Z}_{2}$ symmetries on $\phi_1$ and $\phi_2$ can both be preserved by having $\Phi \to - \Phi$ under either of them.  It is also obvious that integrating out $\Phi$ would not generate any trilinear couplings of the light scalars.  
Note that this model is extremely contrived (or fine-tuned) since loop corrections would tend to generate other interactions.  Here we simply ignore this issue since it is not the main concern of our study.  The main motivation to consider this UV model is that,  
by design, the 4-point amplitude $\A_1 \equiv \A(\phi_1 \phi_1 \phi_1 \phi_1)$ (or $\A_2$) arises only at the one-loop level, so it provides an explicit example for us to apply the dispersion relation on $\A_1$ and study its implication at the one-loop level.    
The full one-loop contribution to $\A_1$ in the UV model contains three different diagrams (each with all possible crossing diagrams, not explicitly shown) as illustrated in \autoref{fig:aphi1}.  Here, the optical theorem implies
\beq
\frac{1}{s} \, {\mathrm{Im}} \left[\A_1|_{t\to 0}\right] = \sigma(\phi_1 \phi_1 \to \phi_2 \phi_2) + \sigma(\phi_1 \phi_1 \to \Phi \Phi) \,, \label{eq:op}
\eeq
where the total cross sections on the right-hand side are computed at the tree level, corresponding to ``cutting and folding'' the one loop elastic amplitudes.  

After integrating out the heavy scalar $\Phi$, we obtain an effective theory with $\phi_1$ and $\phi_2$, which are illustrated on the right panel of \autoref{fig:aphi1}.  The two diagrams involving $\Phi$ in the UV model each  generates both tree-level and one-loop diagrams in the EFT. Correspondingly, the dim-8 operator 
coefficient $c_{1}^{[8]}$ (corresponding to the $(\partial \phi_1)^4$ term) is generated via one-loop matching and it also receives RG mixing contributions from the coefficients of other dim-8 or dim-6 operators which are generated at the tree level.

We note here that, even without the explicit results of matching and dispersion relations, it is clear from \autoref{fig:aphi1} that the anomalous dimension matrices $\gamma_{ij}$ in \autoref{eq:beta}, which correspond to the mixing between different dim-8 coefficients, should not be subject to any positivity bound.  This is because it corresponds, via the optical theorem in \autoref{eq:op}, to the contribution to $\sigma(\phi_1\phi_1 \to \phi_2 \phi_2)$ from the interference term between the diagram with a $t$-channel $\Phi$ and the diagram with the $\lambda$ coupling, which does not have to be positive.   
Indeed, this interference term is proportional to $\lambda g^2$, where $\lambda$ is a renormalizable coupling that could take either sign.


\section{The positivity bounds}
\label{sec:pos}

Positivity bounds can be derived from the dispersion relation of a 4-point forward elastic amplitude multiplied by some function of $s$.  For an elastic amplitude $\A(ab \rightarrow ab)$, the forward amplitude (which we denote as $\tilde{\A}(s)$) is obtained by simply taking the $t\to 0$ limit, which becomes a function of $s$ only.
The original derivation~\cite{Adams:2006sv} requires a small mass gap for the contour to be extended to the entire complex plane.  While it can still be used in our case assuming the light scalars have masses that can be smoothly deformed to zero, it turns out to be more convenient to used a modified version that exploits the crossing symmetry of a real scalar~\cite{Herrero-Valea:2020wxz,Chala:2023jyx}.  Another advantage of this modified version is that it produces a bound that is less sensitive to the scale at which the amplitude is expanded around (which we denote as $s_0$ instead of the usual $\mu^2$, as we will reserve $\mu$ for the renormalization scale), which will be clear in a moment.  
Other approaches are also possible, for instance the {\it arc} variable defined in Ref.~\cite{Bellazzini:2020cot}.  Here we focus on the one in Refs.~\cite{Herrero-Valea:2020wxz,Chala:2023jyx} and leave a detailed comparison of different approaches to future studies.  

\begin{figure}[tbp]
	\centering
	\includegraphics[width=0.55\textwidth]{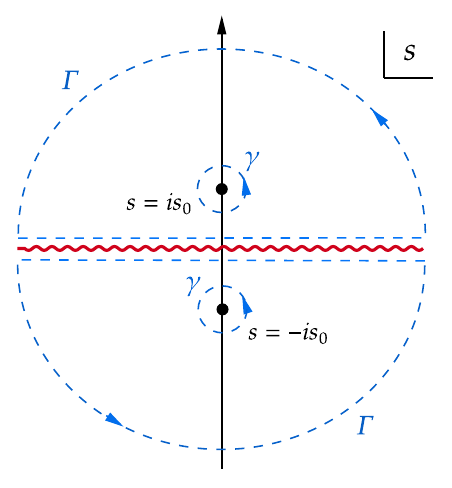}
	\caption{Contours in the complex $s$-plane. The zigzag line represents the branch cut on the real axis. The contours $\gamma$ is equivalent to the contours $\Gamma$, where the radius of the big semi-circles of $\Gamma$ is at $|s| \rightarrow \infty$.} 
    \label{fig:contour}
\end{figure}

The following derivation relies on the crossing symmetry of the amplitude (that it is invariant under the $s \leftrightarrow u$ exchange).  In the $t\to 0$ limit, the massless relation $s+t+u=0$ implies $u=-s$, and crossing symmetry implies that $\tilde{\mathcal{A}}(s)=\tilde{\mathcal{A}}( - s )$.
Performing an analytical continuation of $s$ to the whole complex plane, the real axis contains simple poles and branch cuts which correspond to intermediate particles at the tree and loop levels.  For massless intermediate particles, the branch cuts extends down to the origin and covers the entire real axis, as illustrated in \autoref{fig:contour}.  
Now consider the contour integral 
\begin{equation*}
\oint\limits_{ s = i s_{0} }\!\! \frac{ds}{ 2\pi i } \frac{ s^{3} \tilde{\mathcal{A}} (s) }{ (s^{2} + s_{0}^{2})^{3} }
\end{equation*}
around the point $s=i s_0$ with $s_0>0$.  Due to the crossing symmetry ($s\to -s$), it is equivalent to the same integral around the point $s= - i s_0$.  We now simply add the two contour integrals together and define  
\begin{equation}
\label{eq:Sigma1}
    \Sigma \equiv \oint_{\gamma} \frac{ds}{ 2\pi i } \frac{ s^{3} \tilde{\mathcal{A}} (s) }{ (s^{2} + s_{0}^{2})^{3} } = \Bigg(\ \oint\limits_{ s = i s_{0} } + \oint\limits_{ s = - i s_{0} } \Bigg) \frac{ds}{ 2\pi i } \frac{ s^{3} \tilde{\mathcal{A}} (s) }{ (s^{2} + s_{0}^{2})^{3} }\, ,
\end{equation}
where $\gamma$ indicates the sum of the contours around $s=i s_0$ and $s= - i s_0$.    
We then deform the contours from $\gamma$ to $\Gamma$, as shown in \autoref{fig:contour}, and the contribution along the two big semi-circles vanish due to the Froissart bound, which states that $|\tilde{\mathcal{A}}(s)| < {\rm const} \cdot s \ln^{2} s$ at $s \to \infty$~\cite{Froissart:1961ux,Martin:1962rt}.   
\autoref{eq:Sigma1} could then be written as the difference between the line integral above and below the real axis: 
\begin{equation}
    \begin{split}
    \label{eq:Sigma2}
        \Sigma &=\int_{-\infty}^{\infty} \frac{ds}{2\pi i} \frac{ s^{3} \left( \tilde{\mathcal{A}}(s+i\epsilon) - \tilde{\mathcal{A}}(s-i\epsilon) \right) }{ \left(s^{2} + s_{0}^{2}\right)^{3} } \\
        &=\int_{0}^{\infty} \frac{ds}{\pi i} \frac{ s^{3} \left( \tilde{\mathcal{A}}(s+i\epsilon) - \tilde{\mathcal{A}}(s-i\epsilon) \right) }{ \left(s^{2} + s_{0}^{2}\right)^{3} } \\
        &= \int_{0}^{\infty} \frac{ds}{\pi i} \frac{ s^{3} \left( \tilde{\mathcal{A}}(s+i\epsilon) - \tilde{\mathcal{A}}^{\ast}(s+i\epsilon) \right) }{ \left(s^{2} + s_{0}^{2}\right)^{3} } \\
        &= \frac{2}{\pi} \int_{0}^{\infty} ds\, \frac{s^{3}\, {\mathrm{Im}} \big[\tilde{A}(s)\big] }{ \left(s^{2} + s_{0}^{2}\right)^{3} } \,, 
    \end{split}
\end{equation}
in which $\epsilon$ should be taken infinitesimal. In the second line the $s \leftrightarrow -s$ crossing symmetry is used, and in the third line we have used $\tilde{\mathcal{A}}^{\ast}(s) = \tilde{\mathcal{A}}(s^{\ast})$.  
Applying the optical theorem (as in \autoref{eq:op}), we obtain the dispersion relation
\begin{equation}\label{eq:Sigmapos}
    \Sigma =  \oint_{\gamma} \frac{ds}{ 2\pi i } \frac{ s^{3} \tilde{\mathcal{A}} (s) }{ \left(s^{2} + s_{0}^{2}\right)^{3} }  =  \frac{2}{\pi} \int_{0}^{\infty} ds\, \frac{ s^{4} \sigma(s) }{ \left(s^{2} + s_{0}^{2}\right)^{3} } \geq 0 \,,
\end{equation}
where $\sigma(s)$ is the total cross section, and its positivity implies that $\Sigma \geq 0$.
One nice feature of \autoref{eq:Sigmapos} is that $\Sigma \geq 0$ holds for any value of $s_0$ as long as it is real.  However, for a valid EFT interpretation we would require $s_0 \ll \Lambda^2$, so the left-hand side of \autoref{eq:Sigmapos} can be computed in the EFT with a truncated series.\footnote{More explicitly, the main contribution to $\Sigma$ comes from operators with dimension 8 or lower, while the contribution from dim-10 operators are suppressed by an additional factor of $s_0/\Lambda^2$.}
The right-hand side of \autoref{eq:Sigmapos} can only be computed in the full UV theory, so \autoref{eq:Sigmapos} can be interpreted as a relation between the EFT and the UV physics.   
At the tree level, the (massless) forward amplitude can be written as a polynomial of $s$, $\A(s) = c_{a} s^{a}$ where $a=0,1,2,\ldots$ and $c_{a}$ is a linear combination of the Wilson coefficients of dim-$n$ operators $c^{[n]}_i$ with $n=4+2a$.  By considering the $s_0 \to 0$ limit, \autoref{eq:Sigmapos} then implies $c_2 \geq 0$.  At loop level, however, setting $s_0 \to 0$ introduces log divergences, as we will see later.  It is thus more desirable to have a small but finite $s_0$.


\subsection{Implications on the EFT}
\label{sec:poseft}

We now apply the dispersion relation in \autoref{eq:Sigmapos} to the 2-scalar EFT to study its implications.
A crucial observation here is that \autoref{eq:Sigmapos} is obtained from the optical theorem and is thus valid order by order in the coupling expansion.  As such, one could consider the contribution at a fixed order, and \autoref{eq:Sigmapos} holds as long as all contributions at that order are included on both sides of the equation.  For a tree-level amplitude (which corresponds to $2\to 1$ cross sections via the optical theorem), it is straightforward to compute $\Sigma$, and for the 3 amplitudes in \autoref{tab:eft} we obtain the familiar positivity bounds\footnote{Some nontrivial bounds involving also $c_{12,t}^{[8]}$ can be obtained by considering the amplitudes of linear combinations of $\phi_1$ and $\phi_2$.  They are however not particularly relevant for the discussions below.}
\begin{equation}
    \quad c_{1}^{[8]} \geq 0\,, \quad \quad c_{2}^{[8]} \geq 0\,, \quad \quad c_{12,su}^{[8]} \geq 0 \,. 
\end{equation}
However, an implicit assumption here is that these Wilson coefficients are generated at the tree level in the UV model. If the Wilson coefficient is generated only at the one-loop level in the UV model, the situation could become more complicated, as we already illustrated with the $\Phi\phi_1\phi_2$ model in \autoref{sec:mduv}.  As such, it is important to carefully exam the dispersion relation at the one-loop level.  We will keep the results general in this section by focusing on the 2-scalar EFT and compute $\Sigma$ at the one-loop level.  In \autoref{sec:posuv}, we will apply the results to the $\Phi\phi_1\phi_2$ model. 
Since our main interest is the case where the amplitude is generated only at the one-loop level in the UV, we will focus on the amplitude $\A_1 = \A(\phi_1\phi_1\phi_1\phi_1)$.   The results are obviously applicable to $\A_2$ as well.
The amplitude $\A_{12}= \A(\phi_1\phi_2\phi_1\phi_2)$ has a different kinematic structure, but within the EFT there is no qualitative difference in the results. 
The results for $\A_{12}$ are presented in \autoref{app:A12}.

It is straightforward to compute $\A_1$ up to the one loop level.  We implement the $\overline{\rm MS}$ scheme, with amplitudes having explicit dependence on the renormalization scale $\mu$.  The divergent and logarithmic terms can also be computed within the on-shell framework by taking generalized unitarity cuts~\cite{Bern:1994zx,Bern:1994cg,Britto:2004nc,Mastrolia:2009dr,Jiang:2020mhe,EliasMiro:2020tdv,Baratella:2020lzz}. 
We have, for the dim-4 and dim-6 contributions,
\begin{align}    
        \A_1^{[4]} & =  c_{1}^{[4]}+ \frac{ 1 }{ 32 \pi^{2} } \left( \left( c_{1}^{[4]} \right)^{2} + \left( c_{12}^{[4]} \right)^{2} \right) \left( - \log \frac{ -s }{ \mu^{2} } - \log \frac{ -t }{ \mu^{2} } - \log \frac{ -u }{ \mu^{2} } + 6 \right) \,, \label{eq:A41} \\
        \A_1^{[6]} & = \frac{ 1 }{ 16 \pi^{2} } \left( c_{12}^{[4]} c_{12}^{[6]} \right) \left( - s \log \frac{ -s }{ \mu^{2} } - t \log \frac{ -t }{ \mu^{2} } - u \log \frac{ -u }{ \mu^{2} } \right) \label{eq:A61} \,.
\end{align}         
For the dim-8 contribution,  we have $\A^{[8]}_1 =\A^{ [8], \text{tree} }_1 + \A^{ [8], \text{1-loop} }_1$, where
\begin{equation}
        \A^{ [8], \text{tree} }_1 = c_{1}^{[8]} \left( s^{2} + t^{2} + u^{2} \right) \,,
\end{equation}
and
\begin{equation} \label{eq:A82}
\begin{split}
        \A^{ [8], \text{1-loop} }_1 & = \frac{ 1 }{ 16 \pi^{2} } s^{2} \Bigg[ - \log{ \frac{ -s }{ \mu^{2} } } \left( \frac{1}{2} \left( c_{12}^{[6]} \right)^{2} + \frac{2}{3}\, c_{12}^{[4]} c_{12,su}^{[8]} + c_{12}^{[4]} c_{12,t}^{[8]} + \frac{5}{3}\, c_{1}^{[4]} c_{1}^{[8]} \right) \\ 
        & \quad \quad \quad + \left( c_{12}^{[6]} \right)^{2} + \frac{13}{9}\, c_{12}^{[4]} c_{12,su}^{[8]} + 2\, c_{12}^{[4]} c_{12,t}^{[8]} + \frac{31}{9}\, c_{1}^{[4]} c_{1}^{[8]} \Bigg] \\
        & \quad \quad \quad + \quad (\, s \longleftrightarrow t\, ) \quad + \quad ( \,s \longleftrightarrow u \,) \,.
\end{split} 
\end{equation}
We could now take the forward limit ($t\to 0$) and compute $\Sigma$.  A potential issue is that $\A^{[4]}_1$ is divergent in the $t\to 0$ limit due to the $\log{t}$ contribution (while similar terms in $\A^{[6]}_1$ and $\A^{[8]}_1$ are further suppressed by factors of $t$ or $t^2$ and vanish in the $t\to 0$ limit).  However, only the term proportional to $\log s$ or $\log u$ in $\A^{[4]}_1$ would contribute to $\Sigma$.  
In a more rigorous treatment, one could keep a small scalar mass $m$, compute $\Sigma$ and then take the $m\to 0$ limit.  In the end,  
we have\footnote{Note that the $c_{12}^{[4]} c_{12}^{[6]}$ term in \autoref{eq:Si1loop} has an extra factor of $\pi$ in the coefficient.  This comes from the term $i[\log(-i) - \log(i)]=\pi$ which is generated when plugging \autoref{eq:A61} to the left-hand side of \autoref{eq:Sigmapos}. } 

\begin{equation} \label{eq:Si1loop}
    \begin{split}
        \Sigma & = 2 c_{1}^{[8]} + \frac{ 1 }{ 64 \pi^{2} } \frac{ 1 }{ s_{0}^{2} } \left( \left( c_{1}^{[4]} \right)^{2} + \left( c_{12}^{[4]} \right)^{2} \right) + \frac{ 1 }{ 16 \pi^{2} } \frac{ 1 }{ s_{0} } \frac{ 3 \pi }{ 8 }  c_{12}^{[4]} c_{12}^{[6]} + \left( \frac{3}{4} +  \log \frac{ s_{0} }{ \mu^{2} } \right) \beta_{1}^{[8]} \\ 
        & \quad \quad \quad \quad + \frac{ 1 }{ 16 \pi^{2} } \left( 2 \left( c_{12}^{[6]} \right)^{2} + \frac{26}{9}\, c_{12}^{[4]} c_{12,su}^{[8]} + 4\, c_{12}^{[4]} c_{12,t}^{[8]} + \frac{62}{9}\, c_{1}^{[4]} c_{1}^{[8]} \right) \,,
    \end{split}
\end{equation}
where, since the logarithmic term from $\A^{ [8], \text{1-loop} }_1$ is directly related to the corresponding $\beta$-function $\beta_{1}^{[8]}$, we have conveniently replaced the combination of coefficients by $\beta_{1}^{[8]}$ (see \autoref{eq:beta8}), 
\beq
\beta_{1}^{[8]} = -\frac{ 1 }{ 16 \pi^{2} } \left(  \frac{4}{3}\,  c_{12}^{[4]} c_{12,su}^{[8]} + 2\, c_{12}^{[4]} c_{12,t}^{[8]} + \frac{10}{3}\, c_{1}^{[4]} c_{1}^{[8]} + \left( c_{12}^{[6]} \right)^{2} \right). \label{eq:b81}
\eeq
On the other hand, there is no explicit $\log \left(s_0 / \mu^2\right)$ contribution from $\A^{[4]}_1$ or $\A^{[6]}_1$.  Note that the renormalization scale $\mu$ in \autoref{eq:Si1loop} should be understood as the scale at which all the couplings are defined (and the requirement of amplitudes being independent of $\mu$ leads to the RG running of the couplings).

A few important remarks are in order.  First of all, it is peculiar that $\A^{[4]}_1$ and $\A^{[6]}_1$ have contributions to $\Sigma$ at the one-loop level (which are proportional to $1/s^2_0$ and $1/s_0$, respectively), as they obviously have no contribution at the tree level.  However, this is expected from the optical theorem. 
 In particular, since $c^{[4]}_1$ is a renormalizable coupling that contributes to the tree-level cross section $\sigma(\phi_1\phi_1 \to \phi_1 \phi_1)$, we could check its contribution to $\Sigma$ from the right-hand side of \autoref{eq:Sigmapos},  
\beq
\sigma(\phi_1 \phi_1 \to \phi_1 \phi_1)\big|_{c^{[4]}_1} = \frac{(c^{[4]}_1)^2}{32\pi s} \, \hspace{0.5cm}  \Longrightarrow \hspace{0.6cm}
\Sigma = \frac{2}{\pi} \int_0^{\infty} ds\,\frac{s^4 \sigma}{\left(s^2+s_0^2\right)^3} = \frac{(c^{[4]}_1)^2}{64\pi^2 s_0^2} \,, \label{eq:c41dis}
\eeq
which agrees exactly with \autoref{eq:Si1loop}. The same also holds for $c^{[4]}_{12}$ which contributes to $\sigma(\phi_1\phi_1 \to \phi_2 \phi_2)$.  These contributions are important --- if they are not included, one could easily take the $s_0\to 0$ limit, in which case $\Sigma$ is dominated by the $\beta^{[8]}_1$ contribution, and applying $\Sigma\geq 0$ leads to the incorrect bound $\beta^{[8]}_1 \leq 0$.  
%


With the contributions from $\A^{[4]}_1$ and $\A^{[6]}_1$, it is unclear from \autoref{eq:Si1loop} whether $c^{[8]}_1\geq 0$ still holds even if it is generated at the tree level.  Nevertheless, the tree-level dispersion relation which gives us $c^{[8]}_1\geq 0$ holds in this case, as mentioned above.  
In addition, if $c^{[8]}_1$ is generated at the tree level, then it should not be related to the renormalizable quartic couplings $c^{[4]}_1$ and $c^{[4]}_{12}$, and one could consider the limit where $c^{[4]}_1=c^{[4]}_{12}=0$, and $c^{[8]}_1 \geq 0$ generally holds.  However, {\it a priori} we do not know whether $c^{[8]}_1$ is generated at the tree or loop level, so strictly speaking the statement $c^{[8]}_1\geq 0$ may not be true in the most general case, as already pointed out in Ref.~\cite{Chala:2021wpj}.     
However, if $\phi_{1,2}$ are Nambu-Goldstone bosons (as considered in {\it e.g.} Ref.~\cite{Adams:2006sv, Bellazzini:2020cot}), $c^{[4]}_1$, $c^{[4]}_{12}$ and $c^{[6]}_{12}$ would all vanish, then \autoref{eq:Si1loop} becomes $\Sigma = 2 c^{[8]}_1$, giving a robust positivity bound $c^{[8]}_1 \geq 0$.   
One could also consider the case $c^{[4]}_1=c^{[4]}_{12}=0$ but $c^{[6]}_{12}$ is nonzero,  
and take the $s_0\to 0$ limit.  It is then possible to deduce the bound $\beta^{[8]}_1 \leq 0$, which is however trivially satisfied since $\beta_{1}^{[8]} = - ( c_{12}^{[6]})^{2} /16\pi^2$ in this case.  

Finally, it is interesting to keep $s_0$ finite and take the limit $\mu\to 0$.  This corresponds to the IR limit in the framework of RG flow, where $\log \mu^2$ becomes divergent and a resummation of the leading log terms is required.  Indeed, while it would look like $\Sigma$ is again dominated by the $\beta^{[8]}_1$ term in this limit (now with a positive coefficient), the couplings in \autoref{eq:Si1loop} are also expected to diverge in the $\mu \to 0$ limit without resummation,\footnote{Note that, like the amplitudes, $\Sigma$ is also independent of $\mu$, so the divergence must cancel in the $\mu\to 0$ limit.} so no meaningful positivity bound can be obtained from \autoref{eq:Si1loop} in this limit.  On the other hand, while the resummation of leading log terms is automatically done for the running of couplings by solving the one-loop RG equations (RGEs), it is far less clear how it can be done for amplitudes, since one also needs to obtain the correct kinematic dependence.  It is unclear to us how to obtain a resummed version of \autoref{eq:Si1loop}.  

We also note that, since the cross section can be computed in the EFT, it is possible to subtract a contribution
\begin{equation*}
\frac{2}{\pi} \int_0^{\epsilon \Lambda} \frac{s^4 \sigma}{\left(s^2+s_0^2\right)^3} \,, \quad \quad  \text{with} \quad  \epsilon \lesssim 1 
\end{equation*}
from both sides of \autoref{eq:Sigmapos} to obtain a so-called {\it improved positivity bound}~\cite{Bellazzini:2017fep, deRham:2017xox, Bi:2019phv}, 
\beq
\Sigma' \equiv \Sigma - \frac{2}{\pi} \int_0^{\epsilon \Lambda} \frac{s^4 \sigma}{\left(s^2+s_0^2\right)^3} = \frac{2}{\pi} \int_{\epsilon \Lambda}^\infty \frac{s^4 \sigma}{\left(s^2+s_0^2\right)^3}  \geq 0  \,.
\eeq
Similar results can also be obtained from the {\it arc} variable in Ref.~\cite{Bellazzini:2020cot}. However, for small $\epsilon \Lambda$, $c^{[4]}_1$ and $c^{[4]}_{12}$ still have large contributions to $\Sigma'$ (which are of the same order to the ones in \autoref{eq:c41dis} if $s_0\sim \epsilon \Lambda$), and it is not clear if the positivity of $\Sigma'$ provides additional useful information than the one of $\Sigma$.  We leave a more detailed implementation of the improved positivity bound to future studies.


\subsection{Top-down perspective from the $\Phi\phi_1\phi_2$ model}
\label{sec:posuv}

Having discussed the dispersion relation of $\A_1$ in the EFT, we now move on to the $\Phi\phi_1\phi_2$ model introduced in \autoref{sec:mduv}, and explicitly check the dispersion relation at the one loop level, as illustrated in \autoref{fig:aphi1}.  
The first step is to integrate out $\Phi$ and match the model to the 2-scalar EFT.  Once we obtain the Wilson coefficients at the matching scale, we could then run them down to a lower scale and plug them in \autoref{eq:Si1loop} to compute $\Sigma$.
The matching is performed using the \texttt{Matchete} package~\cite{Fuentes-Martin:2022jrf}, which is based on functional method (see also {\it e.g.} Refs.~\cite{Henning:2014wua,Henning:2016lyp,Cohen:2020fcu}). 
After calculating the one-light-particle-irreducible (1LPI) effective action $\Gamma_{ \text{L} }[\phi]$, the Wilson coefficients are determined by the following matching condition
\begin{equation}
    \Gamma_{ \text{L,EFT} }\left(c_{i}^{[j]}, \mu = \mu_m\right) = \Gamma_{ \text{L,UV} }\left(g,\lambda,\mu=\mu_m\right).
\end{equation}
where $ \mu_m$ is the matching scale. Note there is an underlying assumption here that the UV theory is weakly coupled around the matching scale.

To compute $\Sigma$ to the one-loop order in the UV model, it is important to keep track of two types of contributions.  The first are the Wilson coefficients that contribute to $\Sigma$ at the tree level, and the matching to these coefficients needs to be done to the one-loop level.  According to \autoref{eq:Si1loop}, the only such coefficient is $c^{[8]}_1$, and  
\beq
c_{1}^{[8]}(M) = \frac{ 1 }{ 16 \pi^{2} } \frac{g^2}{M^{4}}  \frac{1}{45} \left( 55 \lambda  - 166 g^{2} \right) \,, \label{eq:c81mat}
\eeq
where we have chosen the matching scale to be $M$ in order to eliminate the log terms.  
At this point, we could already see that $c_{1}^{[8]}(M)$ may take either sign depending on the values of the parameters $\lambda$ and $g$.  
The second type of contributions are those that enter $\Sigma$ at the one-loop level, and those only need to be matched at the tree level.  The non-zero ones are (again at the matching scale $M$)
\beq \label{eq:ctreemat}
\begin{split}
c_{12}^{[4]}(M) &= 2 g^{2} - \lambda \,, \\
c_{12}^{[6]}(M) &= - \frac{ g^{2} }{ M^{2} } \,, \\
c_{12,su}^{[8]}(M) &= \frac{  g^{2} }{ M^{4} } \,, 
\end{split}
\eeq
where $c_{12}^{[4]}$ also receives a contribution proportional to $g^2$ from tree-level exchanges of $\Phi$.  
We then run these Wilson coefficients down to scale $\mu$ using the $\beta$-functions in \autoref{eq:beta8} and \autoref{eq:beta6} and plug them in \autoref{eq:Si1loop}.  Note that, without resummation, the solutions of the RGEs have the general form
\begin{equation}\label{eq:RGsol}
    c^{[n]}_{i}(\mu) = c_{i}^{[n]}(M) + \beta_{i}^{[n]}(M) \log \frac{ \mu }{ M }.
\end{equation}
When plugged in $\A_1$ or $\Sigma$, the $\mu$-dependent terms cancel as intended.  All-in-all, the final result is
\begin{align}    
        \Sigma & = \frac{ \lambda^{2} }{ 64 \pi^{2} } \frac{ 1 }{ s_{0}^{2} } + \frac{ \lambda g^{2} }{ 16 \pi^{2} } \left( - \frac{ 1 }{ s_{0}^{2} } + \frac{ 3 \pi }{ 8 } \frac{ 1 }{ M^{2} s_{0} } + \frac{5}{9} \frac{ 1 }{ M^{4} } + \frac{4}{3} \frac{ 1 }{ M^{4} } \log \frac{ s_{0} }{ M^{2} } \right) \nonumber \\
        & \quad \quad + \frac{ g^{4} }{ 16 \pi^{2} } \left( \frac{ 1 }{ s_{0}^{2} } - \frac{ 3 \pi }{ 4 } \frac{ 1 }{ M^{2} s_{0} } - \frac{47}{20} \frac{1}{M^{4}} - \frac{11}{3} \frac{ 1 }{ M^{4} } \log \frac{ s_{0} }{ M^{2} } \right)  \,, \label{eq:Sig1UV} 
\end{align}
where the couplings $\lambda$ and $g$ are defined at the matching scale $M$.  It is also informative to have the explicit form of $\beta_{1}^{[8]}$, which is
\beq
\beta_{1}^{[8]} =\frac{1}{16 \pi^2}\left(\frac{4}{3} \frac{\lambda g^2}{M^4}-\frac{11}{3} \frac{g^4}{M^4}\right) \,. \label{eq:b81uv}
\eeq

With an explicit UV model, $\Sigma$ can also be computed from the total cross section using the right-hand side of \autoref{eq:Sigmapos}, which provides an important check.  Here, the total cross section is given by the sum of the two tree-level 2-to-2 cross sections $\sigma(\phi_1 \phi_1 \to \phi_2 \phi_2)$ and $\sigma(\phi_1 \phi_1 \to \Phi \Phi)$.  It is straightforward to compute these two cross sections, which are
\beq
\begin{split} 
\sigma(\phi_{1} \phi_{1} \rightarrow \phi_2 \phi_2) & = \frac{\lambda^2}{32 \pi s} - \frac{\lambda g^2 M^2}{8 \pi s^2}  \log \left(1+\frac{s}{M^2}\right) \\
& \quad \quad + \frac{g^4 M^2}{16 \pi s}\left[\frac{1}{s+M^2} + \frac{2 M^2}{s \left(s+2 M^2 \right)} \log \left(1+\frac{s}{M^2}\right) \right],
\end{split}
\eeq
and 
\beq 
    \sigma(\phi_{1} \phi_{1} \rightarrow \Phi \Phi) = \frac{ g^{4} }{ 16 \pi s } \sqrt{ 1 - \frac{4M^2}{s} }
     \left\{ 1 + \frac{ 2 M^4 \log \left[\frac{s - 2M^2 + \sqrt{s \left(s-4M^2 \right)}}{s - 2M^2 - \sqrt{s \left(s-4M^2 \right)}}\right] }{ \left( s-2 M^2 \right) \sqrt{s \left(s-4M^2 \right)}}  \right\} \Theta( s - 4M^2 ) \,,
\eeq
where $\Theta( s - 4 M^{2} )$ is the Heaviside step function.
Plugging the cross sections into the right-hand side of \autoref{eq:Sigmapos} and expanding in terms of $1/M^2$, we obtain
\begin{align}
 \frac{2}{\pi} & \int_{0}^{\infty} ds\, \frac{s^{4} \sigma(\phi_{1} \phi_{1} \rightarrow \phi_2 \phi_2) }{ \left(s^{2} + s_{0}^{2}\right)^{3} }  = \frac{ \lambda^{2} }{ 64 \pi^{2} } \frac{ 1 }{ s_{0}^{2} } + \frac{ \lambda g^{2} }{ 16 \pi^{2} } \left( - \frac{ 1 }{ s_{0}^{2} } + \frac{ 3 \pi }{ 8 } \frac{ 1 }{ M^{2} s_{0} } + \frac{5}{9} \frac{ 1 }{ M^{4} } + \frac{4}{3} \frac{ 1 }{ M^{4} } \log \frac{ s_{0} }{ M^{2} } \right) \nonumber \\
& + \frac{g^4}{16 \pi^2}\left[ \frac{1}{s_0^2} - \frac{3 \pi}{4} \frac{1}{M^2 s_0}-\left( \frac{\pi^2}{16}+\frac{16}{9} \right) \frac{1}{M^4} - \frac{11}{3} \frac{1}{M^4} \log \frac{s_0}{M^2}\right] + \mathcal{O}\left(\frac{s_0}{M^6}\right) \,,
\end{align}
and
\begin{align}
& \frac{2}{\pi} \int_{0}^{\infty} ds\, \frac{s^{4} \sigma(\phi_{1} \phi_{1} \rightarrow \Phi \Phi) }{ \left(s^{2} + s_{0}^{2}\right)^{3} } ~= ~ \frac{g^4}{128\pi^2 M^4} \left( \frac{\pi^2}{2} - \frac{206}{45} \right) + \mathcal{O}\left(\frac{s_0}{M^6}\right) \,.
\end{align}
Adding the two contributions together, the result indeed agrees with \autoref{eq:Sig1UV}.  It should be noted that, in the computation of the one-loop amplitude (the $\tilde{\mathcal{A}} (s)$ on the left-hand side of \autoref{eq:Sigmapos}), a regularization-renormalization procedure has been performed, with the divergence cancelled by counter terms, while no such procedure has been done for the right-hand side of \autoref{eq:Sigmapos}.  However, in our case the one-loop dim-8 contribution is finite since the corresponding one-loop counter term cannot be generated in the UV model.  On the other hand, $\A_1$ contains a dim-4 counter term (since the $\phi^4_1$ interaction is not forbidden but only tuned to zero), 
but it would not contribute to $\Sigma$.\footnote{The contributions of the contour terms have the same kinematics as the corresponding tree-level ones, so only dim-8 (or higher) contour terms contribute to $\Sigma$.}  It is possible that, beyond one-loop, or in a more general case, the dim-8 counter term would contribute to $\Sigma$, and a regularization-renormalization procedure would also be needed for the right-hand side of \autoref{eq:Sigmapos} in order to have a meaningful comparison.  The details of such a procedure are beyond the scope of this paper.

$\Sigma$ has three different contributions which are proportional to $\lambda^2$, $\lambda g^2$ and $g^4$, respectively.  They correspond to the three rows in \autoref{fig:aphi1}. One important observation, as already mentioned in \autoref{sec:mduv}, is that the term proportional to $\lambda g^2$ corresponds to an interference contribution to the total cross section on the right-hand side of \autoref{eq:Sigmapos}, and could take either sign.  It can be clearly seen in the $\Phi\phi_1\phi_2$ model since the renormalizable coupling $\lambda$ could take either sign.  Furthermore, while the requirement of a stable vacuum could impose non-trivial bounds on $\lambda$, we note here that a positive $\lambda$, corresponding to a positive potential at larger field values, gives a negative $\lambda g^2$ term in $\Sigma$ when $s_0 \ll M^2$, as shown in \autoref{eq:Sig1UV}.  Correspondingly, $\beta^{[8]}_1$ also has a contribution that is proportional to $\lambda g^2$.   
One interesting limit to consider here is that $\lambda \gg g^2$, in which case one could keep only the $\mathcal{O}(\lambda^2)$ and $\mathcal{O}(\lambda g^2)$ contributions, and omit the $\mathcal{O}(g^4)$ ones.  In this case, there is no positivity bound whatsoever on $c^{[8]}_1$ or $\beta^{[8]}_1$, since they are both proportional to $\lambda g^2$.

It is also illustrative to consider the $\lambda \to 0$ limit.  The terms proportional to $\lambda^2$ or $\lambda g^2$ then vanishes in \autoref{eq:Sig1UV}, and we have
\beq
\Sigma|_{\lambda\to 0} = \frac{ g^{4} }{ 16 \pi^{2} } \left( \frac{ 1 }{ s_{0}^{2} } - \frac{ 3 \pi }{ 4 } \frac{ 1 }{ M^{2} s_{0} } - \frac{47}{20} \frac{1}{M^{4}} - \frac{11}{3} \frac{ 1 }{ M^{4} } \log \frac{ s_{0} }{ M^{2} } \right)  \,, \label{eq:Sig2UV}
\eeq
which turns out to be negative at $s_0=M^2$, seemingly violating the positivity bound.  However, as we have emphasized, $\Sigma$ is only strictly positive if all the one-loop contributions are included, and in particular the ones from higher dimensional operators (dim-10 and above) which are all important at $s_0=M^2$.  Indeed, we see that for the region $s_0 \ll M^2$ where the EFT is valid, $\Sigma$ is clearly positive as expected.\footnote{More precisely, \autoref{eq:Sig2UV} becomes positive for $s_0\lesssim 0.47 M^2$.} 
On the other hand, $c^{[8]}_1$ is also negative at the matching scale $\mu=M$, as shown in \autoref{eq:c81mat}.\footnote{A similar example in the SMEFT framework was also observed in Ref.~\cite{Chala:2021wpj}, where the $\beta$-function also drives the dim-8 coefficient positive at lower scales.}  The $\beta$-function in \autoref{eq:b81uv} is also negative, and only at a sufficiently low scale ($\mu^2 \lesssim 0.13 M^2$)
will $c^{[8]}_1$ change sign and become positive.

\section{Conclusion}
\label{sec:con}

In this paper, we carefully analysed the implication of positivity bounds at the one-loop level in a 2-scalar ($\phi_1,\,\phi_2$) EFT with explicit computations of the $\phi_1\phi_1 \to \phi_1\phi_1$ amplitude and its dispersion relation.  For the positivity bound to hold, it is crucial to include all the contributions on both sides of the dispersion relation.  Different from the tree-level case where only the dim-8 (or even higher dimensional) effects contribute, at the one-loop level there are also contributions from lower dimensional operators.  These contributions play an important role in the interpretation of the positivity bounds in the case where the dim-8 contribution is only generated at the one-loop level in the UV theory.  With all contributions included, we found that the $\beta$-functions (or the anomalous dimension matrix) of the dim-8 coefficients are generally not subject to positivity bounds.  In particular, the RG mixing between different dim-8 coefficients corresponds to an interference contribution to the total cross section under the optical theorem, and could take either sign.    
In special cases, for instance where the dim-4 couplings vanish, a stronger statement could then be made for the signs of the loop-generated dim-8 coefficients or the $\beta$-functions.  
We also verified our results with a UV model involving a heavy scalar $\Phi$ and a $\Phi\phi_1\phi_2$ trilinear coupling, in which the $\phi_1\phi_1 \to \phi_1\phi_1$ amplitude is generated at the one-loop level.  This model provides an explicit example on how the na\"ive tree-level positivity bound appears to be violated when the dim-8 contribution is generated at the one-loop level, and how the bound is ``restored'' if all the one-loop contributions to the dispersion relation are included.

\begin{figure}[t]
	\centering
	\includegraphics[width=0.83\textwidth]{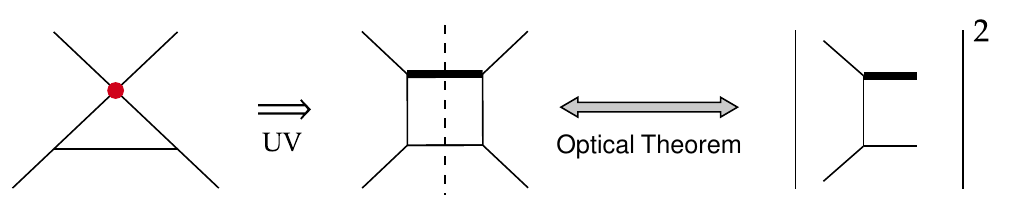}
	\caption{With renormalizable trilinear couplings it is possible to construct a ``symmetrical'' diagram with one insertion of a dim-8 operator, whose UV completion ({\it e.g.} a heavy particle denoted by the thick line) corresponds to a square term of a cross section, and could be subject to a positivity bound.}
    \label{fig:optical2}
\end{figure}

Our study confirms some of the findings in the previous studies  
while also clarifies a number of important points.   
Like the previous studies, it will be desirable to generalize our results to more practical EFTs, in particular the SMEFT.  While the general principles should still apply, the amplitudes in SMEFT involve particles with masses and spins which may require more careful treatments.  In many cases, the amplitudes also contain more propagators (instead of just 4-point contact interactions) which could have nontrivial contributions in the dispersion relation.  For instance, with renormalizable trilinear couplings it is possible to construct a ``symmetrical'' one-loop elastic amplitude with one insertion of a dim-8 operator that could be subject to a positivity bound, as illustrated in \autoref{fig:optical2}.  It will be interesting to check whether the examples in Refs.~\cite{Chala:2023jyx,Chala:2023xjy} belong to this category.  
It will also be interesting to find more examples where the one-loop dim-8 contributions could violate the na\"ive tree-level positivity bound.  
In this case, if an apparent violation of the positivity bound on the dim-8 Wilson coefficient is found by experiments, it may not indicate the break down of QFT, but rather that the tree-level assumption is invalid for the dim-8 contribution.  Furthermore, to test positivity bounds, one needs to measure the interference term between SM and the dim-8 contribution.  This requires a sizable SM amplitude, so a tree-level SM coupling is usually needed, and a large SM contribution to the dispersion relation (as in \autoref{eq:Si1loop}) at the one-loop level is almost guaranteed.    
If such cases generally exist, it would be of crucial importance for the experimental tests of positivity bounds, and additional measures ({\it e.g.} of the energy dependence of the new physics contribution) will be needed to make the test more robust.   
On the other hand, it is interesting that a tree-level UV completion could be ruled out if the experimental results turn out to fall into these scenarios.  
We leave these important topics to be explored by future studies.

%


\section*{Acknowledgments}

We thank Mikael Chala, Xu Li, Chia-Hsien Shen, Lian-Tao Wang and Shuang-Yong Zhou for useful discussions and valuable comments on the manuscripts.  This work is supported by the National Natural Science Foundation of China (NSFC) under grant No. 12035008 and No. 12375091.  


\newpage

\appendix

\section{Full $\beta$-functions and one-loop matching results}
\label{app:beta}

The one-loop $\beta$-functions for the dim-8 coefficients (defined as $\beta_{i}^{[8]} \equiv \mu \frac{ d c_{i}^{[8]} }{ d \mu }$) are
\begin{equation}
\begin{split} \label{eq:beta8}
    \beta_{1}^{[8]} &= -\frac{ 1 }{ 16 \pi^{2} } \left( \frac{4}{3}\, c_{12}^{[4]} c_{12,su}^{[8]} + 2\, c_{12}^{[4]} c_{12,t}^{[8]} + \frac{10}{3}\, c_{1}^{[4]} c_{1}^{[8]} + \left( c_{12}^{[6]} \right)^{2} \right), \\ 
    \beta_{2}^{[8]} &= -\frac{ 1 }{ 16 \pi^{2} } \left( \frac{4}{3}\, c_{12}^{[4]} c_{12,su}^{[8]} + 2\, c_{12}^{[4]} c_{12,t}^{[8]} + \frac{10}{3}\, c_{2}^{[4]} c_{2}^{[8]} + \left( c_{12}^{[6]} \right)^{2} \right), \\ 
    \beta_{12,su}^{[8]} &= -\frac{1}{ 16 \pi^{2} } \left( \frac{16}{3}\, c_{12}^{[4]} c_{12,su}^{[8]} + \frac{4}{3}\, c_{12}^{[4]} c_{12,t}^{[8]} + \frac{2}{3} \left( c_{12}^{[6]} \right)^{2} \right), \\ 
    \beta_{12,t}^{[8]} &= -\frac{ 1 }{ 16 \pi^{2} } \left( \frac{2}{3} \left( c_{1}^{[4]} + c_{2}^{[4]} \right) c_{12, su}^{[8]} + \left( c_{1}^{[4]} + c_{2}^{[4]} \right) c_{12,t}^{[8]}\right. \\
    & \quad \quad \quad \quad \quad \quad \quad \quad \quad \quad \ + \left.\frac{5}{3}\, c_{12}^{[4]} \left( c_{1}^{[8]} + c_{2}^{[8]} \right) - \frac{1}{3} \left( c_{12}^{[6]} \right)^{2} \right). 
\end{split}
\end{equation}
The one-loop $\beta$-functions of dim-4 and dim-6 coefficients are
\begin{equation}
\label{eq:beta6}
\begin{split} 
    \beta_{1}^{[4]} &= -\frac{ 3 }{ 16 \pi^{2} } \left( \left( c_{1}^{[4]} \right)^{2} + \left( c_{12}^{[4]} \right)^{2} \right), \quad\quad \beta_{2}^{[4]} = -\frac{ 3 }{ 16 \pi^{2} } \left( \left( c_{2}^{[4]} \right)^{2} + \left( c_{12}^{[4]} \right)^{2} \right),\\
     \beta_{12}^{[4]} &= -\frac{ 1 }{ 16 \pi^{2} }\, c_{12}^{[4]} \left( c_{1}^{[4]} + c_{2}^{[4]} + 4 c_{12}^{[4]} \right) \,, \quad \quad 
    \beta_{12}^{[6]} = -\frac{ 1 }{ 16 \pi^{2} }\, c_{12}^{[6]} \left( c_{1}^{[4]} + c_{2}^{[4]} + 2 c_{12}^{[4]} \right)\,. 
\end{split}
\end{equation}
The full one-loop matching results for the $\Phi\phi_1\phi_2$ model in \autoref{eq:phiuv} are (with matching scale $\mu=M$)
\begin{equation}
    \begin{split}
        c_{12}^{[4]}(M) & = 2 g^{2} - \lambda - \frac{ 1 }{ 16 \pi^{2} } g^{2} \left( 12 g^{2} - 5 \lambda \right), \quad \quad 
        c_{1}^{[4]}(M) = \frac{ 1 }{ 16 \pi^{2} } g^{2} \left( 6 \lambda - 12 g^{2} \right), \\ 
        c_{2}^{[4]}(M) & = \frac{ 1 }{ 16 \pi^{2} } g^{2} \left( 6 \lambda - 12 g^{2} \right), \quad \quad 
        c_{12}^{[6]}(M) = - \frac{ g^{2} }{ M^{2} } \left[ 1 - \frac{ 1 }{ 16 \pi^{2} } \left( \frac{23}{6} g^{2} + \frac{3}{2} \lambda \right) \right] , \\
        c_{1}^{[8]}(M) & = \frac{ 1 }{ 16 \pi^{2} } \frac{ g^{2} }{M^{4}} \frac{1}{45} \left( 
        55 \lambda - 166 g^{2} \right), \quad  
        c_{2}^{[8]}(M) = \frac{ 1 }{ 16 \pi^{2} } \frac{ g^{2} }{M^{4}} \frac{1}{45} \left( 
        55 \lambda - 166 g^{2} \right), \\
        c_{12,su}^{[8]}(M) & = \frac{ g^{2} }{ M^{4} } \left[ 1 - \frac{ 1 }{ 16 \pi^{2} } \frac{11}{9} \left( 4 g^{2} - \lambda \right) \right], \quad \quad 
        c_{12,t}^{[8]}(M) = \frac{ 1 }{ 16 \pi^{2} } \frac{g^{4}}{M^{4}} \frac{29}{45} \,,
    \end{split}
\end{equation}
which are obtained with the \texttt{Matchete} package~\cite{Fuentes-Martin:2022jrf}.

\section{The results for $\phi_1 \phi_2 \to \phi_1 \phi_2$}
\label{app:A12}

Here we provide the results for the $\phi_1 \phi_2 \to \phi_1 \phi_2$ process in the EFT, including the amplitude $\A_{12} \equiv \A(\phi_1 \phi_2  \phi_1 \phi_2)$ at the one-loop level and the corresponding $\Sigma$ as defined in \autoref{eq:Sigmapos}.  The amplitudes are given by (again with the notation in \autoref{eq:ampcom})

\begin{equation}
    \begin{split}
        \mathcal{A}_{12}^{[4]} =& \frac{ 1 }{ 16 \pi^{2} } \left( c_{12}^{[4]} \right)^{2} \left( - \log \frac{ -s }{ \mu^{2} } - \log \frac{ -u }{ \mu^{2} } + 4 \right) \\ 
        &+ \frac{ 1 }{ 16 \pi^{2} } \left( c_{1}^{[4]} + c_{2}^{[4]} \right) c_{12}^{[4]} \frac{1}{2} \left( - \log \frac{ t }{ \mu^{2} } + 2 \right) + c_{12}^{[4]} \,,
    \end{split}
\end{equation}
\begin{equation}
    \begin{split}
        \mathcal{A}_{12}^{[6]} =& \frac{ 1 }{ 16 \pi^{2} } \left( c_{12}^{[6]} c_{12}^{[4]} \right) \left( s \log \frac{ -s }{ \mu^{2} } + u \log \frac{ -u }{ \mu^{2} } + 2 t \right) \\
        &+ \frac{ 1 }{ 16 \pi^{2} } \left( c_{1}^{[4]} + c_{2}^{[4]} \right) c_{12}^{[6]} \left( - \frac{1}{2} t \log \frac{ -t }{ \mu^{2} } + t \right) + c_{12}^{[6]}\, t \,,
    \end{split}
\end{equation}
and
\begin{align} 
        \mathcal{A}_{12}^{[8],\text{tree}} & = c_{12,su}^{[8]} ( s^{2} + u^{2} ) + c_{12,t}^{[8]} t^{2} \,, 
\end{align} 
\begin{align}
        \mathcal{A}_{12}^{[8],\text{1-loop}} & = \frac{ 1 }{ 16 \pi^{2} } s^{2} \Bigg[ - \log{ \frac{ -s }{ \mu^{2} } } \left( \frac{8}{3}\, c_{12}^{[4]} c_{12,su}^{[8]} + \frac{2}{3}\, c_{12}^{[4]} c_{12,t}^{[8]} + \frac{1}{3} \left( c_{12}^{[6]} \right)^{2} \right)  \nonumber\\ 
        & \quad \quad \quad \quad \quad \quad \quad \quad \quad \quad \quad  
        + \frac{49}{9}\, c_{12}^{[4]} c_{12,su}^{[8]} + \frac{13}{9}\, c_{12}^{[4]} c_{12,t}^{[8]} + \frac{13}{18} \left( c_{12}^{[6]} \right)^{2} \Bigg] \nonumber \\
        & \quad 
        + \frac{ 1 }{ 16 \pi^{2} }\, st \left( c_{12}^{[6]} \right)^{2} \left( - \frac{1}{6} \log \frac{ -s }{ \mu^{2} } + \frac{8}{3} \right) 
           \quad\quad\quad + \quad ( s \longleftrightarrow u ) \nonumber \\
        + \frac{ 1 }{ 16 \pi^{2} } t^{2} & \Bigg[ - \log \frac{ -t }{ \mu^{2} } \left( \frac{1}{3} \left( c_{1}^{[4]} + c_{2}^{[4]} \right) c_{12,su}^{[8]} + \frac{1}{2} \left( c_{1}^{[4]} + c_{1}^{[4]} \right) c_{12,t}^{[8]} +  \frac{5}{6}\, c_{12}^{[4]} \left( c_{1}^{[8]} + c_{2}^{[8]} \right) \right) \nonumber \\ 
        & \quad \quad + \frac{13}{18}\, c_{12,su}^{[8]}\left( c_{1}^{[4]} + c_{2}^{[4]} \right) + c_{12,t}^{[8]} \left( c_{1}^{[4]} + c_{2}^{[4]} \right) + \frac{31}{18} \, c_{12}^{[4]} \left( c_{1}^{[8]} + c_{2}^{[8]} \right) \Bigg] \,. 
\end{align} 
The corresponding $\Sigma$ is given by
\begin{equation}
\label{eq:sigma_1212}
    \begin{split}
        \Sigma & =  2\,c_{12,su}^{[8]}+ \frac{ 1 }{ 32 \pi^{2} } \frac{ 1 }{ s_{0}^{2} } \left( c_{12}^{[4]} \right)^{2} - \frac{ 1 }{ 16 \pi^{2} } \frac{ 1 }{ s_{0} } \frac{ 3 \pi }{ 8 }  c_{12}^{[4]} c_{12}^{[6]} + \left( \frac{3}{4} + \log \frac{ s_{0} }{ \mu^{2} } \right) \beta_{12,su}^{[8]} \\ 
        & \quad \quad \quad + \frac{ 1 }{ 16 \pi^{2} } \left( \frac{98}{9}\, c_{12}^{[4]} c_{12,su}^{[8]} + \frac{ 26 }{ 9 }\, c_{12}^{[4]} c_{12,t}^{[8]} + \frac{13}{9} \left( c_{12}^{[6]} \right)^{2} \right) \,.
    \end{split}
\end{equation}
Note that the one-loop contributions to $\A_{12}^{[8],\text{1-loop}}$ from $c^{[8]}_1 $ and $c^{[8]}_2$ are proportional to $t^2$ and vanish in the forward limit.  As a result, $c^{[8]}_1 $ and $c^{[8]}_2$ do not contribute to $\Sigma$.  This can also be understood from the fact that the corresponding amplitudes could not be cut in a way to give the total cross section term on the right-hand side of \autoref{eq:Sigmapos}.

\bibliographystyle{JHEP}

\bibliography{posloop}

\end{document}